\documentclass[aps,prl,twocolumn,superscriptaddress,showpacs,preprintnumbers,amsmath,amssymb]{revtex4}
\usepackage{graphicx} % Include figure files
\usepackage{dcolumn}  % Align table columns on decimal point
\usepackage{lipsum}

\newcommand{\epm} {e^{+}e^{-}}
\newcommand{\emu} {e\textrm{-}\mu}
\newcommand{\taupm} {\tau^{+}\tau^{-}}
\newcommand{\piKs} { \pi^{-} K^0_S} 
\newcommand{\piKspizero} { \pi^{-} K^0_S \pi^{0} }
\newcommand{\piKsKs} { \pi^{-} K^0_S K^0_S }
\newcommand{\piKsKspizero} { \pi^{-} K^0_S K^0_S \pi^{0} }
\newcommand{\KKs} { K^{-} K^0_S }
\newcommand{\KKspizero} { K^{-} K^0_S \pi^{0} }

\newcommand{\Ks}  { K^{0}_{S} }

\newcommand{\Br}{\mathcal{B}} 
\newcommand{\tauTO} {\tau^{-} \rightarrow }

{}
\graphicspath{{ps}}

\begin{document}

%\vspace*{-3\baselineskip}
%\resizebox{!}{3cm}{\includegraphics{belle.eps}}

\preprint{\vbox{ \hbox{   }
%                 \hbox{BELLE-CONF-12nn}
%                 \hbox{TAU2012-09}
%                 \hbox{hep-ex nnnn, if available}
}}

\title{ \quad\\[0.5cm]  Measurement of the branching fractions and mass spectra 
for $\tau$ lepton decays including $K_{S}^{0}$ at Belle 
\footnote{Proceeding of Tau2012, the 12th international workshop on tau lepton physics Nagoya, Japan, 17-21 September, 2012 }}

\author{S. Ryu}
%%%% >>>>> insert the authorlist here. BEFORE the abstract !!!!! <<<<<
%%%% >>>>> obtain the latest summer conference authorlist from the
%%%% >>>>> authorship confirmation web page

\collaboration{Belle Collaboration}
\noaffiliation

\begin{abstract}
%{\it Author instructions:} This template file is updated in June-2008. \\ 
%To determine the BELLE-CONF number, 
% and ICHEP06 abstract number (to be available), 
%check the Belle internal WEB page for conference papers,
%"http://belle.kek.jp/secured/conferences/ICHEP2008/".
% For publications that have been submitted recently, the only change
% that is required is the inclusion of the BELLE-CONF
% number %, ICHEP2008 abstract numbers 
% on the upper right hand corner of the first page.
%Note that for all conference papers except for those submitted to journals,
%a single common authorlist is used. This is prepared by
%Leo Piilonen and is available on the authorship confirmation WEB page.
%{\bf The conference paper
%will be submitted to the conference and posted on the public WEB page
%by one of the analysis coordinators after approval
%by the internal referees.} Conference papers are intended as precursors
%to publications. This template is designed so that with simple
%modifications, a Belle conference paper
% can easily be converted into a format suitable for publication.
%{\it end of Author instructions}. 

We report a study of $\tau$ lepton decays involving $\Ks$ with 
a 669 fb$^{-1}$ data sample accumulated with the Belle detector 
at the KEKB asymmetric-energy $e^{+}e^{-}$ collider.
The branching fractions have been measured for the 
$\tauTO \piKs \nu_{\tau}$, $\KKs \nu_{\tau}$, $\piKspizero \nu_{\tau}$, 
$\KKspizero \nu_{\tau}$,
$\piKsKs \nu_{\tau}$ and $\piKsKspizero \nu_{\tau}$ decays.
We also provide the unfolded mass spectra for 
$\tauTO \piKspizero \nu_{\tau}$ and $\tauTO \KKspizero \nu_{\tau}$.
\end{abstract}

%\pacs{13.65.+i, 13.25.Gv, 14.40.Gx}

\maketitle

%%%% >>>> keep the final version single-spaced
\tighten

\section{Introduction}
\label{sec:introduction}
\newdimen\origiwspc
\newdimen\origiwstr
\origiwspc=\fontdimen2=\fontdimen2\font

%\fontdimen2\font=\origiwspc
$\tau$ lepton is the only known lepton massive enough to decay 
into hadrons. Its hadronic decays are ideally suited to investigate the 
hadronic weak currents.
Former studies for semileptonic $\tau$ decays done at LEP and 
CLEO~\cite{ALEPHOPALCLEO} provide accurate results for measurements 
of branching fractions as well as spectral functions for one- or 
three-prong decay modes with any number of neutral mesons that do not 
suffer from Cabibbo and/or phase space suppression.
However, because of limited statistics, their studies for 
Cabibbo-suppressed or multi-prong $\tau$ decay modes do not provide 
sufficient information for investigating the hadronic structure 
or for testing important standard model parameters, e.g., $|V_{us}|$.

In this analysis we use a data sample of 669 fb$^{-1}$ corresponding 
to 616$\times 10^6$ $\taupm$ pair events, which is two orders of magnitude 
larger than those that were available prior to the $B$-factory experiments.
The $\taupm$ pair events, accumulated with the Belle detector~\cite{Belle}, 
are produced at the KEKB asymmetric-energy $\epm$ collider~\cite{KEKB} 
running in the energy range of the $\Upsilon(4S)$ with a cross section 
comparable to that of $B\bar{B}$~\cite{Banerjee}.
It is also worth noting that the $\tau^{+}\tau^{-}$ produced at the 
$\Upsilon(4S)$ are in a very clean experimental environment 
with low background.

The leptonic decay modes are used here for tagging $\tau$ events 
for a precise measurement of the four decay modes: $\tauTO \piKs \nu_{\tau}$, 
$\tauTO \KKs \nu_{\tau}$, $\tauTO \piKspizero \nu_{\tau}$ and 
$\tauTO \KKspizero \nu_{\tau}$.
The other two decay modes, $\tauTO \piKsKs \nu_{\tau}$ and 
$\tauTO \piKsKspizero \nu_{\tau}$, however, are selected using 
one-prong decays for tagging to increase the signal.

\section{Event selection} 
\label{sec:eventselection}
The selection of $\taupm$ events is performed in two stages.
At the first stage, loose conditions are applied for the primary 
$\taupm$ selection. 
A $\taupm$ event produced back-to-back in the $\epm$ center-of-mass (CM) 
frame is divided into two hemispheres 
according to the plane perpendicular to the thrust axis.
Each event is required to have only one track on one side of the 
hemisphere (tag side) and any number of tracks on the other side (signal side). 

As a second step, we classified the events into two types, leptonic and hadronic events, using the combination of the number of charged tracks and 
their flavors on each side.
The events on both sides containing a single track are classified 
as candidates of leptonic events, while the events where the tag side 
contains a single track with three or more tracks on the signal side 
are classified as hadronic events.

Particle identification (PID) is used to determine the flavor of charged 
tracks on both sides by defining and using the likelihood variables to 
identify the electron and muon track on the tag side and the charged 
pion or kaon on the signal side except the $\Ks$ daughter tracks.

The leptonic events in which the $\tau$ lepton on one side decays 
to $e^{-} \bar{\nu}_{e} \nu_{\tau}$ and the other side to  
$\mu^{-}\bar{\nu}_{\mu}\nu_{\tau}$ ($\emu$ events), are selected for 
normalization. 
The branching fractions for $\tau^{-} \to e^{-}\bar{\nu}_{e}\nu_{\tau}$ 
and $\tau^{-} \to \mu^{-}\bar{\nu}_{\mu}\nu_{\tau}$ are measured with 
good precision and can be used as a normalization for the 
branching fraction measurement. 
The detection efficiency and its statistical error for $\emu$ events
is (19.31 $\pm$ 0.03)\%.
 
Hadronic events are selected by different tagging with at least 
one $\Ks \to \pi^{+}\pi^{-}$, which is reconstructed with a vertex fit 
using the momentum of two oppositely charged tracks on the signal side.
The lepton-tagged events are defined as those with one $\tau$ decaying 
to leptons, while the other decays to hadrons. 
Lepton tagging is used for $\piKs$, $\KKs$, 
$\piKspizero$ and $\KKspizero$ events. 
Since the signal yield for events involving more than one $\Ks$ is 
smaller than that involving a single $\Ks$, we use both lepton and 
hadron tagging for $\piKsKs$ and $\piKsKspizero$ events. 
The decay modes used for hadron tagging are 
$\tau^{-} \to h^{-}(\geq 0\ \pi^{0}) \nu_{\tau}$ where $h = \pi, K$.

The signal $\pi^{0}\rightarrow\gamma\gamma$ is reconstructed from 
the invariant mass determined from the momenta of two photons detected 
on the signal side. 
The distribution of the difference between the invariant mass of 
the two photons and the nominal $\pi^0$ mass normalized to the resolution, 
$S_{\gamma\gamma} = (m_{\gamma\gamma}-m_{\pi^{0}})/\sigma_{\gamma\gamma}$, is used 
to determine the number of genuine $\pi^{0}$'s and to estimate the 
level of background from mass sidebands. 

\section{Determination of branching fractions}
\label{sec:exclusivebr}
The branching fractions for decay modes containing a single $\Ks$ 
are obtained using
\begin{eqnarray}
\mathcal{B}(\tau^{-} \to X^{-}\nu_{\tau}) = \frac{N^{\rm Sig}_{X \textrm{-} l }}{N_{e \textrm{-}\mu}} \frac{\mathcal{B}_{e} \mathcal{B}_{\mu}}{\mathcal{B}_{e}+\mathcal{B}_{\mu}},
\label{eq:bfltag}
\end{eqnarray}
where $X$ is the signal channel under study, 
$N^{\rm Sig}_{X \textrm{-} l} \equiv N_{X \textrm{-} l}/\epsilon_{X \textrm{-} l}$ 
is the number of events with an efficiency correction, where the $\tau$ 
lepton on one side decays into a signal mode and the $\tau$ lepton on the 
other decays into leptons,
$N_{e \textrm{-} \mu} \equiv N^{\rm obs}_{e \textrm{-}\mu}/\epsilon_{e \textrm{-}\mu}$ 
is the number of events with an efficiency correction for $\emu$ events,
$\mathcal{B}_{e}$ and $\mathcal{B}_{\mu}$ are the world-average branching 
fractions 
for $\tauTO e^{-}\bar{\nu}_{e}\nu_{\tau}$ and 
$\tauTO \mu^{-}\bar{\nu}_{\mu}\nu_{\tau}$~\cite{PDG}, respectively.

In order to increase the number of candidates, $\piKsKs$ and 
$\piKsKspizero$ events are selected by both lepton and hadron tagging.
To determine the branching fraction for multiple $\Ks$ modes, we used 
\begin{eqnarray}
\mathcal{B}(\tau^{-} \to X^{-}\nu_{\tau}) = \frac{N^{\rm Sig}_{X \textrm{-} A}}{ 2 \times N_{\tau\bar{\tau}} \times \mathcal{B}(\tauTO A^{-}\bar{\nu}\nu)},
\label{eq:bfatag}
\end{eqnarray}
where $A$ indicates a one-prong combination of particles: {\it e.g.}, 
$\tauTO l^{-}\bar{\nu}_{l}\nu_{\tau}$ and 
$\tauTO h^{-}(\geq 0\ \pi^{0})\nu_{\tau}$.

In both cases, the number of signal events $N^{\rm Sig}_{i}$ is given as
\begin{eqnarray}
N^{\rm Sig}_{i} = \sum_{j} (\mathcal{E}^{-1})_{ij}(N^{\rm Data}_{j} - N^{\rm Bg}_{j}),
\label{eq:nsigexp}
\end{eqnarray}
where $i$ is the true decay channel of interest and $j$ is the decay channel 
reconstructed in the analysis. $N^{\rm Data}_{i}$ is the number of selected 
events and $N^{Bg}_{j}$ is the estimated background coming from 
decay modes other than six under consideration, and non-$\tau$ processes.
$\mathcal{E}_{ij}$ is the efficiency matrix.
The diagonal components are the efficiency for the $i$-th channel and 
the other components are the probability of migration from the 
$i$-th to $j$-th decay modes.

For the branching fraction measurement, the number of the background events 
coming from $\tau$ decays other than the six modes studied should be determined.
To determine the background, generic MC samples of $\tau$ lepton decays
generated using branching fractions provided by the Particle Data Group 
(PDG) are used.
The non-$\tau$ decay contributions are relatively small, at a level of 1\%, 
and dominated by $q\bar{q}$ continuum events.

\section{Efficiency and corrections}
 \label{sec:efficiency}
In order to determine the signal efficiency and cross-feed rates 
for the decay modes of interest, a 17.2~million $\taupm$ MC sample is used.
Since a difference between MC and data is inevitable, we evaluate 
corrections associated with a difference between MC and data for PID and 
the reconstruction of the neutral particles, $\Ks$ and $\pi^{0}$. 
The corrections of the PID efficiency for electron and muon are derived 
from two-photon events and those for charged pions and kaons are derived 
from $D^{*-} \to D^{0} \pi^{-}$ and $D^{0} \to K^{-} \pi^{+}$ control samples.
The correction for the $\Ks$ reconstruction efficiency obtained is 
0.979 $\pm$ 0.007 using $D^{*-} \rightarrow D^0\pi^{-}$, 
$D^{0}\rightarrow \Ks \pi^{+}\pi^{-}$ control samples.
The MC-data correction for the $\pi^{0}$ efficiency is determined to be 
$0.957 \pm 0.015$.

For the $\piKspizero$ and $\KKspizero$ modes, the number of cross-feed 
events from $\piKs$ and $\KKs$ with spurious $\pi^{0}$ is estimated 
and subtracted by using sidebands of $S_{\gamma\gamma}$.
By this subtraction, the cross-feed contributions from 
$\tauTO \piKs \nu_{\tau}$ and/or $\tauTO \KKs \nu_{\tau}$ are removed 
successfully.

By taking into account the corrections and the efficiency losses by 
subtraction, the efficiency migration matrix $\mathcal{E}_{ij}$ is obtained 
and summarized in Table~\ref{tab:effmatrix}. 
The diagonal elements represent the signal efficiency, otherwise - 
the cross-feed rates among the decay modes of interest.

\begin{table}[t]
\scriptsize
\begin{tabular*}{1.0\columnwidth} {l @{} c @{} c @{} c @{} c @{} c @{} c } \hline
 & \multicolumn{6}{c}{Efficiency matrix(\%)} \\
Reconst.    & \multicolumn{6}{c}{Generated decay mode} \\ 
decay mode   & $\piKs$ \ & $\KKs$ \ & $\piKspizero$ \ & $\KKspizero$ \ & $\piKsKs$ \ & $\piKsKspizero$ \\ \hline
        $\piKs$             & 7.09 & 1.65 & 1.07 & 0.31 & 0.67 & 0.13 \\
        $\KKs$              & 0.35 & 6.69 & 0.06 & 1.01 & 0.04 & 0.01 \\
        $\piKspizero$       & 1.0e-4& 2.6e-3 & 2.65 & 0.54 & 0.51 & 0.23 \\
        $\KKspizero$        & 4.0e-4& 1.0e-4 & 0.11 & 2.19 & 0.01 & 0.00 \\
        $\piKsKs$           &       & 1.0e-4 & 1.0e-4 &     & 2.47 & 0.53 \\
        $\piKsKspizero$     &     &     &    &    & 0.04 & 0.81 \\ \hline

\end{tabular*}
\caption{Efficiency migration matrix $\mathcal{E}_{ij}$ for six decay channels.
The first four rows show the efficiency for lepton tagging, while the 
last two rows show the one for lepton and hadron tagging.}
\label{tab:effmatrix}
\end{table}

\begin{table}[t]
\footnotesize
\begin{tabular*}{1.0\columnwidth}{ l @{~} c @{~} c @{~} c @{~} c @{~} c @{~} c } \hline
             & \multicolumn{6}{c}{$\bigtriangleup \Br/\Br$ (\%)}  \\ 
source & $\piKs$ & $\KKs$ & $\piKspizero$ & $\KKspizero$  & $\piKsKs$ & $\piKsKspizero$ \\ \hline
\ EFF     & 2.7 & 3.5 & 3.2 & 3.3 & 3.4 & 5.6 \\
\ HDM       & -   & -   & 0.3 & 3.4 & -   & -   \\
\ BGE              & 0.2 & 0.3 & 1.9 & 0.4 & 1.8 & 3.2 \\ 
\ NORM  					& 0.4 & 0.5 & 0.4 & 0.5 & 1.3 & 1.3 \\ 
\ $\gamma$ veto            & 0.1 & 1.8 & 1.2 & 1.5 & 1.0 & 2.0 \\ 
\ Total                    & 2.8 & 3.9 & 4.0 & 5.0 & 4.2 & 6.9 \\ \hline
\end{tabular*}
\caption{Summary of the systematic errors.}
\label{tab:systematicseffmatrix} 
\end{table}

\begin{table}[t]
\scriptsize
\begin{tabular*}{1.02\columnwidth}{l @{~} c @{~} c @{~} c @{~} c @{~} c @{~} c} \hline 
               & $\piKs$ & $\KKs$  & $\piKspizero$ & $\KKspizero$  & $\piKsKs$ & $\piKsKspizero$ \\  \hline
$\piKs$        &    1    & -0.28   & -0.08   &  0.01 & -0.01  &  3.9e-3  \\
$\KKs$         &         &    1    &  0.03   & -0.14 & -0.9e-3&  0.3e-3 \\ 
$\piKspizero$  &         &         &    1    & -0.21 & -0.06  &  0.02 \\ 
$\KKspizero$   &         &         &         &   1   & 2.9e-3 & -1.0e-3 \\
$\piKsKs$      &         &         &         &       &    1   & -0.37 \\ 
$\piKsKspizero$&         &         &         &       &        &    1  \\ \hline
\end{tabular*}
\caption{Coefficients of the covariance matrix for statistical 
$\bigoplus$ systematic uncertainty covariance.}
\label{tab:totalcovariancematrix}
\end{table}

\begin{figure}
\includegraphics[width=0.95\columnwidth]{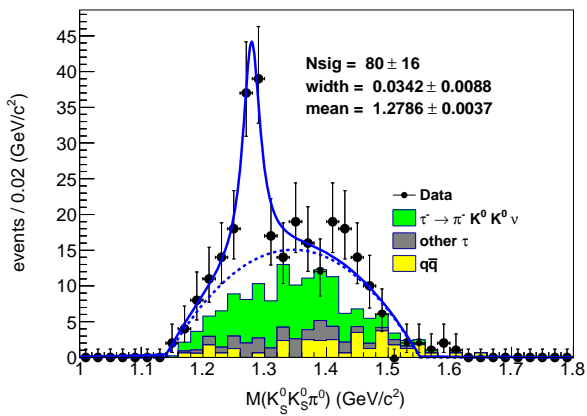}
\caption{Invariant mass of $\Ks\Ks\pi^{0}$ in 
$\tauTO \piKsKspizero \nu_{\tau}$ events. 
A significant $f_{1}(1285)$ signal is seen. The solid line is a fit 
with a Breit-Wigner and a second-order polynomial function.}
\label{fig:mkskspi0}
\end{figure}

\section{Systematic uncertainty}
\label{sec:systematics}
The sources of systematic uncertainties are categorized into the 
uncertainty of detection/reconstruction efficiency(EFF), hadron 
decay models (HDM), background estimation (BGE), normalization (NORM) 
and rejection of energetic photons ($\gamma$ veto).
The uncertainty of efficiency consists of several items: track finding, 
PID, $\Ks$ and $\pi^{0}$ reconstruction and the $\pi^{0}$ sideband 
subtraction for $\piKspizero$ and $\KKspizero$.
These uncertainties are integrated into the efficiency migration 
matrix~\cite{LEFEBVRE2000}.
The uncertainties from the decay model implemented into the $\tau$ 
MC sample are estimated by using weighting sets of MC samples generated 
with different hadron decay models. 
The background uncertainty mostly contributed by other $\tau$ lepton 
decays is estimated by varying the fractions according to their 
uncertainties in PDG. 
By adding these uncertainties in quadrature, 
the total systematic uncertainties for $\piKs$, $\KKs$, $\piKspizero$, 
$\KKspizero$, $\piKsKs$ and $\piKsKspizero$ are 
2.8\%, 3.9\%, 4.0\%, 5.0\%, 4.2\% and 6.9\%, respectively.
The summary of systematic uncertainties is shown in 
Table~\ref{tab:systematicseffmatrix}.

The correlations of these uncertainties are also taken into account 
by the covariance matrix cov($\Br_{i},\Br_{j}$) using the formula 
given in Ref.~\cite{LEFEBVRE2000}. The results including both systematic 
and statistical uncertainties are shown as the coefficients of the 
covariance ${\rm cov}(\Br_{i},\Br_{j})/\sqrt{{\rm cov}(\Br_{i},\Br_{i})~{\rm cov}(\Br_{j},\Br_{j})}$ in Table~\ref{tab:totalcovariancematrix}.

\section{Results}
Using the formula given in Eq.~(\ref{eq:bfltag}), Eq.~(\ref{eq:bfatag}) 
and Eq.~(\ref{eq:nsigexp}), the branching fractions for six decay modes 
are obtained simultaneously.
The branching fraction for $\tauTO \piKs \nu_{\tau}$ is found to be 
$(4.13\pm0.01\pm0.12) \times 10^{-3}$, which is consistent with our 
previous study~\cite{EPIFANOV2007}.
The branching fractions for $\tau^{-} \to \KKs \nu_{\tau}$,
$\piKspizero \nu_{\tau}$ and $\KKspizero \nu_{\tau}$ are measured to 
be $(7.36\pm0.04\pm0.29) \times 10^{-4}$, 
$(1.92\pm0.02\pm0.08) \times 10^{-3}$, and 
$(7.44\pm 0.11 \pm 0.37) \times 10^{-4}$, respectively, 
which are among the most precise results to date.
%In particular, the branching fratio for the $\piKsKspizero$ channel is not determined but an upper limit, $\Br(\tauTO \piKsKspizero \nu_{\tau}) < 2.0\times 10^{-4}$ has been given~\cite{PDG}. 
The branching fractions for $\piKsKs$ and $\piKsKspizero$ have been 
measured for the first time to be $(2.39\pm0.03\pm0.09) \times 10^{-4}$ 
and $(2.06\pm 0.13\pm0.14) \times 10^{-5}$, respectively.

\begin{figure*}
\centering
\includegraphics[width=\columnwidth]{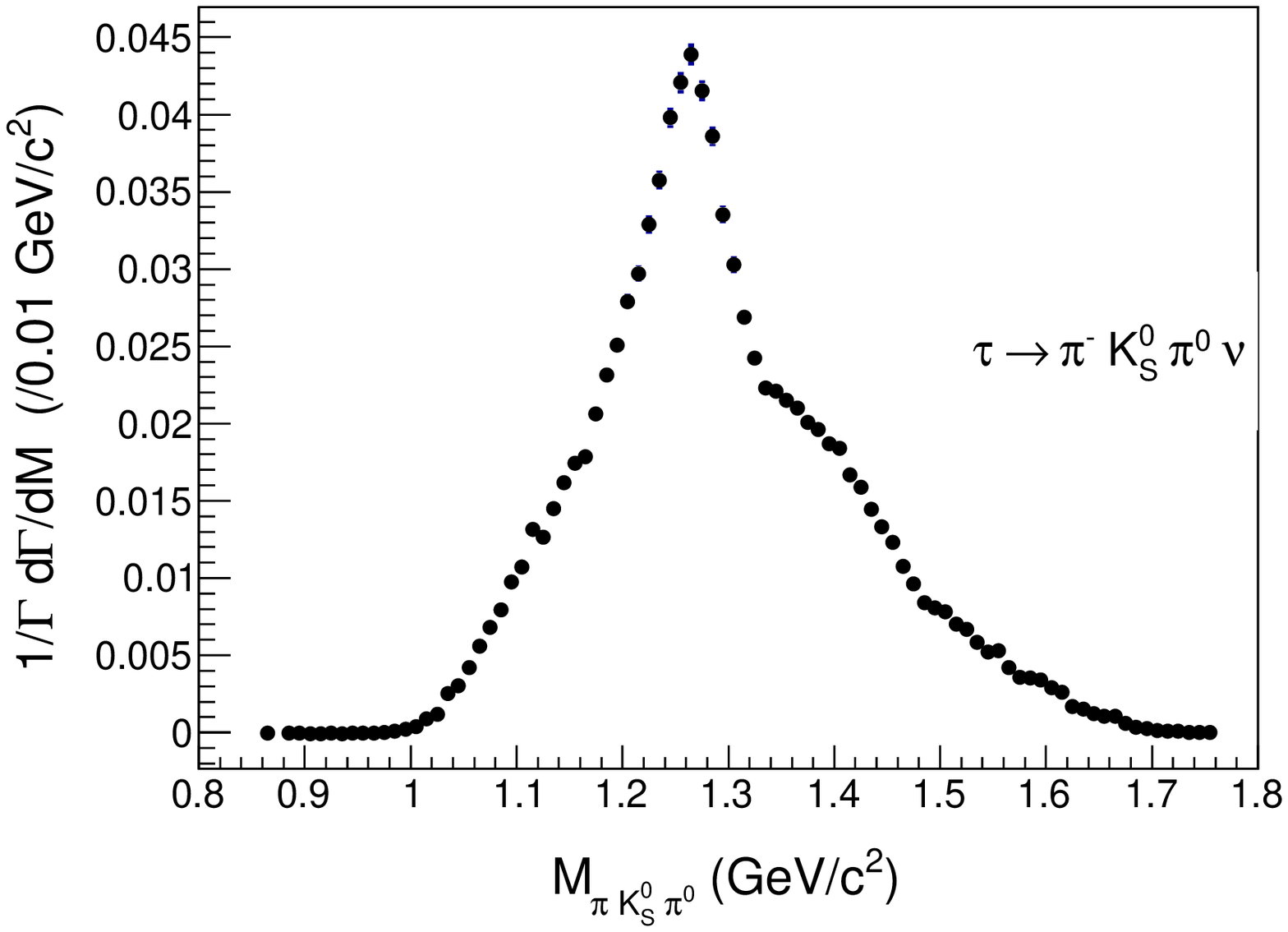}
\includegraphics[width=\columnwidth]{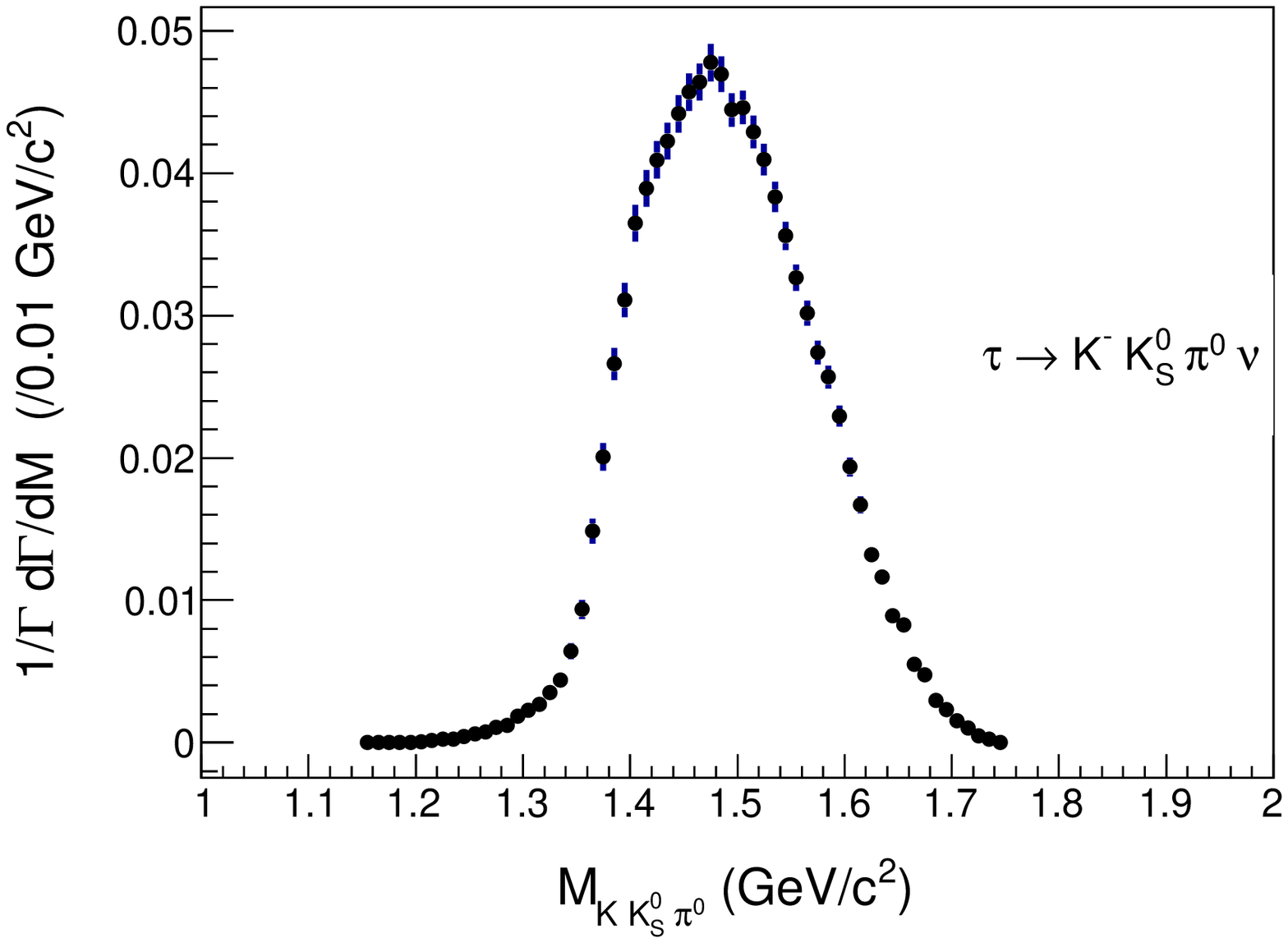}
\caption{Unfolded invariant mass spectra for $\piKspizero$(left) in 
$\tauTO \piKspizero \nu_{\tau}$ and 
$\KKspizero$(right) in $\tauTO \KKspizero \nu_{\tau}$.
The error bars show statistical error and are within the closed
circles in many points.}
\label{fig:unfoldedspectra}
\end{figure*}

\section{Study of $\tauTO \piKsKspizero \nu_{\tau}$}
Not only the branching fraction but also the hadron current 
can be studied in $\tauTO \piKsKspizero \nu_{\tau}$ using the invariant 
mass of $\Ks\Ks\pi^{0}$, $M(\Ks\Ks\pi^{0})$. 
As can be seen in Fig.~\ref{fig:mkskspi0}, the $M(\Ks\Ks\pi^{0})$ 
distribution shows a significant peak at around 1280 MeV/$c^{2}$, 
which can be understood as the $f_{1}(1285)$ resonance in $\tau$ decays.
In order to determine the mass and width for the resonances, 
we used the probability distribution function (PDF) consisting of a 
Breit-Wigner function for the signal and a second-order polynomial 
for the background.
So the signal is parameterized as
\begin{equation}
S(m) = \frac{mM\Gamma_{0}}{ (M^{2}-m^{2})^{2} + (m\Gamma_{0})^{2}}
\end{equation}
where $M$ and $\Gamma_{0}$ are the nominal mass and width, $m$ is the 
mass of the $\Ks\Ks\pi^{0}$ system. 
The total number of signal events is found to be 80$\pm$16. 
The mass and width are 1278$\pm$4 MeV/$c^{2}$ and 35$\pm$9 MeV/$c^{2}$,
consistent with the $f_{1}(1285)$ properties.
%is present in $\tauTO \piKsKspizero \nu_{\tau}$ events as an intermediate hadronic current. 
From the total number of $\tauTO \piKsKspizero \nu_{\tau}$ and the number 
of $f_{1}(1285) \to \Ks\Ks\pi^{0}$ among the 
$\tauTO \piKsKspizero \nu_{\tau}$ events, the branching fraction 
of $\tauTO f_{1}(1285) \pi^{-} \nu_{\tau} \to \piKsKspizero \nu_{\tau}$ 
is found to be $(1.05 \pm 0.24)\times 10^{-5}$.
In this fit, possible interference between $f_{1}$ and non-$f_{1}$ amplitude is ignored,
which will cause considerable model uncertainties in the $f_{1}$ parameters and the
branching fraction. 
Further studies are in progress.

\section{Unfolded mass spectra for $\tauTO \piKspizero \nu_{\tau}$ 
and $\tauTO \KKspizero \nu_{\tau}$}
The mass spectra provide important information for hadronic weak currents:
the $\piKspizero$ mode contributes to the Cabibbo-suppressed weak 
current, while the $\KKspizero$ mode has both vector and axial-vector 
components. The axial-vector component is interesting as the 
Wess-Zumino-Witten anomalous term~\cite{SCHEK} .
The measured mass spectra are distorted due to the finite resolution 
and the efficiency of the detectors. In addition, there is a sizable 
cross-feed between these two modes.
By using the singular value decomposition (SVD) unfolding technique, 
the total mass and submass spectra are obtained from 
$\tauTO \piKspizero \nu_{\tau}$ and $\tauTO \KKspizero \nu_{\tau}$ 
by taking into account the effects from the detector efficiencies, 
resolutions as well as the cross-feed backgrounds.
The results are shown in Figs.~\ref{fig:unfoldedspectra}.
A clear resonance-like structure due to $K_1$(1280) 
is seen in $\tauTO \piKspizero \nu_{\tau}$.

\section{Conclusion}
\label{sec:conclusion}
Using $616 \times 10^{6}$ $\taupm$ events colllected with the Belle detector,
% together with the $e\textrm{-}\mu$ normalization methods and efficiency migration matrix, 
we measured several branching fractions for hadronic $\tau$ decays 
involving $\Ks$: $\piKs$, $\KKs$, $\piKspizero$, $\KKspizero$, 
$\piKsKs$ and $\piKsKspizero$ and provided the covariance matrix, 
which includes both 
statistical and systematic uncertainties.
We also provide the correlation of the systematic errors as the covariance 
matrix.
The invariant mass of $\Ks\Ks\pi^{0}$ in the $\piKsKspizero$ mode shows 
a clear peak around the mass of 1280 $\textrm{MeV}/c^{2}$, which most 
probably comes from the $f_{1}(1285)$ current.
In addition, we measured the unfolded mass spectra for 
$\tauTO \piKspizero \nu_{\tau}$ and $\tauTO \KKspizero \nu_{\tau}$ using the 
SVD unfolding technique.

%\begin{table}[htb]
%\caption{ This is an example of a table in the style preferred by
%Physical Review.}
%\label{sys1}
%\begin{tabular}
%{@{\hspace{0.5cm}}l@{\hspace{0.5cm}}|@{\hspace{0.5cm}}c@{\hspace{0.5cm}}}
%\hline \hline
%Source & Systematic error (\%) \\
%\hline
%ISR correction & $\pm 19$ \\
%Fitting procedure & $\pm 16$ \\
%$J/\psi$ polarization & $\pm 11$ \\
%File handling & $\pm 4$ \\
%Missing runs & $\pm 2$ \\
%Track reconstruction  & $\pm 5$ \\
%Lepton identification  & $\pm 3$ \\
%\hline
%Total & $\pm 28$ \\
%\hline \hline
%\end{tabular}
%\end{table}

%{\it See http://belle.kek.jp/secured/publication/figure\_tips.html for
%advice and macros for publication quality figures.}
%{\bf Please try to make data points and fit
%curves clear and visible. Use reasonable bin sizes and appropriate
%aspect ratios. Axes should be labeled. For color figures
%use primary colors and avoid "Miami Vice" pastels (pink, yellow, 
%light green, etc.).}
%{\it Files for figures should be collected together in a single directory.
%The filenames for figures should correspond to the numbering in the
%paper e.g. psiks\_fig1.eps, psiks\_fig2.eps, psiks\_fig3.eps. These
%figures will be made available to conference speakers and reviewers.
\section*{}
%----------- Long version, for most papers ----------- 
We thank the KEKB group for the excellent operation of the
accelerator, the KEK cryogenics group for the efficient
operation of the solenoid, and the KEK computer group and
the National Institute of Informatics for valuable computing
and SINET3 network support. We acknowledge support from
the Ministry of Education, Culture, Sports, Science, and
Technology of Japan and the Japan Society for the Promotion
of Science; the Australian Research Council and the
Australian Department of Education, Science and Training;
the National Natural Science Foundation of China under
contract No.~10575109 and 10775142; the Department of
Science and Technology of India; 
the BK21 program of the Ministry of Education of Korea, 
the CHEP SRC program and Basic Research program 
(grant No.~R01-2005-000-10089-0) of the Korea Science and
Engineering Foundation, and the Pure Basic Research Group 
program of the Korea Research Foundation; 
the Polish State Committee for Scientific Research; 
%-> remove for now: under contract No.~2P03B 01324; 
the Ministry of Education and Science of the Russian
Federation and the Russian Federal Agency for Atomic Energy;
the Slovenian Research Agency;  the Swiss
National Science Foundation; the National Science Council
and the Ministry of Education of Taiwan; and the U.S.\
Department of Energy.

\end{document}